\documentclass{emulateapj}
\usepackage{apjfonts}

\def\msol{\hbox{$\rm\thinspace M_{\odot}$}} 
\def\etal{{\it et al.\thinspace}}  
\def\eg{{\it e.g.\ }}  

\def\ie{{\it i.e.\ }}

\def\p3m{P${}^3$M}  
\def\ap3m{AP${}^3$M}  
\def\-{{\em{---}}}
  
\def\msun{{M_\odot}}
\def\lesssim{\mathrel{\hbox{\rlap{\hbox{\lower4pt\hbox{$\sim$}}}\hbox{$<$}}}}
\def\gtrsim{\mathrel{\hbox{\rlap{\hbox{\lower4pt\hbox{$\sim$}}}\hbox{$>$}}}}

\newcommand{\be}{\begin{equation}} \newcommand{\ba}{\begin{eqnarray}}
\newcommand{\ee}{\end{equation}} \newcommand{\ea}{\end{eqnarray}}

\begin{document}

\title{Predictions of Quasar Clustering: Redshift, Luminosity and
Selection Dependence}

\author{Robert J. Thacker\altaffilmark{1,2}, Evan
Scannapieco\altaffilmark{3}, H. M. P. Couchman\altaffilmark{4}
and Mark Richardson\altaffilmark{1}}

\altaffiltext{1}{Department of  Astronomy and Physics, Saint  Mary's
University, Halifax, Nova Scotia, B3H 3C3, Canada.}
\altaffiltext{2}{Canada Research Chair}   
\altaffiltext{3}{School of Earth and Space Exploration,  Arizona State 
University, PO Box 871404, Tempe, AZ, 85287-1404.}   
\altaffiltext{4}{Department of Physics and
Astronomy, McMaster University, 1280 Main St.\ West, Hamilton,
Ontario, L8S 4M1, Canada.}

\begin{abstract}
We show that current clustering observations of quasars and luminous
AGN can be explained by a merger model augmented by feedback from
outflows.  Using numerical simulations large enough to study
clustering out to 25 comoving $h^{-1}$ Mpc, we calculate  correlation
functions, biases, and correlation lengths as a function of AGN
redshift and optical and X-ray luminosity.   At optical wavelengths,
our results match a wide range of current observations and generate
predictions for future data sets.  We reproduce the weak
luminosity  dependence of clustering over the currently well-measured
range, and predict a much stronger dependence at higher
luminosities.   The  increase in the amplitude of binary quasar
clustering observed in the Sloan Digital Sky Survey (SDSS) is also
reproduced and is predicted to occur at higher redshift, an effect
that is due to the one halo term in the correlation function.  On the
other hand, our results do not match the rapid evolution  of the
correlation length observed in the SDSS at $z\simeq 3$, a discrepancy that
is at least partially due to differences in the scales probed by
our simulation versus this survey.  In fact, we show that changing the
distances sampled from our simulations can produce changes as
large as 40\% in the fitted correlation lengths.  Finally, in the
X-ray, our simulations produce correlation lengths similar to that
observed in the {\em Chandra} Deep Field (CDF) North, but not the
significantly larger correlation length observed in the CDF South.
\end{abstract}


\section{Introduction}

Despite numerous observational efforts, quasar clustering and its
dependency on luminosity remains controversial.
Early studies suggesting  that clustering decreases with
redshift  (Iovino \& Shaver 1988; Croom \& Shanks 1996) are opposed by
more recent observations, which suggest a more complicated clustering
history that gradually increases with redshift
(Kundic 1997; La  Franca  \etal 1998; Porcani \etal 2004; Croom \etal
2005).   In particular, $z < 2 $ studies with the Sloan Digital Sky
Survey  (SDSS, Myers  \etal 2006,
2007a, 2007b) and 2QZ  (Croom \etal 2005, Porcani \& Norberg 2006)
have uncovered weak evidence for clustering evolution,  
while above $z=2$ the SDSS indicates
strong evolution in clustering  (Shen \etal 2007) and
2QZ shows a somewhat weaker increase 
(Croom \etal 2005).   Over all these redshifts, the luminosity
dependence of clustering is weak (Porciani \etal 2004;
Adelberger \& Steidel 2005; Croom \etal 2005; Myers  2006, 2007a),
which is usually interpreted as being problematic for quasar
models in which AGN luminosity is correlated with proxies for
halo mass.

While questions of obscuration and completeness surround optical
selection techniques, hard X-ray observations are largely
unaffected by obscuration, thus making them perhaps the best candidate for
identifying AGNs (Mushotsky 2004).  At redshifts $z<1$ considerable
effort has been put into measuring X-ray selected AGNs in both hard 
(2-10 keV)  and soft (0.5-2 keV) bands 
(\eg Mullis \etal 2004; Gilli \etal 2005;  
Basilakos
\etal 2004, 2005; Yang \etal 2006; Miyaji \etal 2007). While optical 
$z < 1$ surveys tend to produce correlations lengths between 5-6
$h^{-1}$ Mpc (e.g. Porciani \& Nordberg 2006), X-ray selected catalogs at
these redshifts give
correlation lengths in the range 7-8 $h^{-1}$ Mpc (\eg Mullis \etal
2004). However, the X-ray selected  AGN from the 
{\em Chandra} Deep Field-North and South (CDF-N and CDF-S), exhibit
significant variances in both the correlation length and power-law
slope fits.   Thus, Gilli \etal (2005), using the full catalog from 0.5-8
keV, found a correlation length of  $5.5 \pm 0.6$ $h^{-1}$ Mpc for
the CDF-N, in agreement with optical surveys, and  $10.3 \pm 1.7$
$h^{-1}$ Mpc for the CDF-S, which is clearly in disagreement. This
is surprising given that the $\log{N}-\log{S}$ of the two fields agree well.
A recent re-analysis of this data, binned by luminosity and separating 
into hard and soft classifications, has yielded essentially the same 
overall result (Plionis \etal 2008), but elucidated that the two fields 
have much more consistent clustering behavior when binned by luminosity. 
Previous analyses had suggested the difference in clustering could be 
attributed solely to sample variance due to the lack of large 
superclusters in the CDF-S (Gilli \etal 2003).

While AGN/quasar feedback has become theoretically favored as a
necessary component of galaxy evolution (\eg Scannapieco \& Oh 2004, 
hereafter SO04; Granato \etal 2004; Croton \etal 2006),
direct observational evidence of this idea remains somewhat weak. The
paucity of high-redshift X-ray objects also makes studying clustering
evolution of X-ray catalogs difficult. However, Francke \etal (2008) have
cross-correlated AGN from the Extended CDF-S and luminous blue galaxy
sources at $z \approx 3$ identified in the MUSYC survey (Gawiser \etal
2006). Their results indicate that the AGN targeted in the survey are
more clustered than star-forming luminous blue galaxies, a result consistent with
the idea that typical AGNs tend
to sit in more massive halos than the average galaxy population. Clearly
there remain many open questions about the clustering properties of
optical and X-ray selected AGN, and the ongoing CDF controversy suggests
that we can learn a great deal by comparing to predictions of clustering
from simulations.

From a theoretical perspective, early models of quasar formation
associated quasars  with galaxy mergers and assumed a close
relationship between black hole mass and luminosity (\eg Kauffmann \&
Haehnelt 2000;  Wyithe \& Loeb 2002, 2003).  In this case, the black
hole mass was calculated  either by using the $M_{\rm BH}-\sigma$
relationship (Ferrarese \& Merrit 2000) or associating $M_{\rm BH}$
with the halo circular velocity (Merrit \& Ferrarese 2001; Tremaine
\etal 2002; Ferrarese 2002). Such ``light bulb'' models  successfully
match the luminosity function of high-redshift quasars (\eg Wyithe \&
Loeb 2003), but become progressively more inaccurate at low redshifts
when feedback processes become important (\eg SO04). 
Cosmological simulations using the $M_{\rm
BH}-\sigma$ framework and incorporating feedback have managed to
reproduce the turn-down in the quasar luminosity function with
moderate success (Thacker \etal 2006, hereafter TSC06). However, there
appear to be differences between the detailed behavior of gas in
simulations versus semi-analytic models, which are primarily due to
differences between shock-heating in a uniform medium  relative to an
inhomogeneous one (\eg Helley \etal 2003; Nagamine \etal 2005; Croton
\etal 2006;  Cattaneo \etal 2006; Cattaneo \etal 2007; Di Matteo \etal
2008).

More recent models (\eg Hopkins \etal 2005a, 2005b, 2005c,  2006a,
2006b, 2007a), motivated by numerical modeling of black hole accretion
during mergers (Di Matteo \etal 2005; Springel \etal 2005a, 2005b;
Robertson \etal 2006a, 2006b; Cox \etal 2006a, 2006b), suggest that 
quasar activity is comparatively decoupled from galaxy mass.   This 
picture entails complex relationships between a distinct sequence of AGN
evolutionary  epochs and the feedback processes that regulate them
(\eg Hopkins \etal 2007a).  The resulting behavior is one in which the
bright end of the luminosity function corresponds to quasars radiating
at close to their peak luminosities near the Eddington limit,
while the faint end corresponds to
the same population radiating in the faint part  of their
light curve, at or below $\approx$ 0.1 of the Eddington luminosity
(Hopkins \etal 2005b). As a result, clustering is only a  weak function of
luminosity  (Lidz \etal 2006). While the exact dynamics of nuclear
accretion  flows are still  beyond the resolution of simulations of
colliding galaxies, and are  still the subject of  much active
research and modeling (\eg Proga \etal 2008), the overall
phenomenology in this model of quasar activity is well understood.
Recent increases in computing capacity have lead to simulations of
this model in a cosmological environment and investigations of the
impact of AGN on disk  formation  (\eg Sijacki  \etal 2007; Di Matteo
\etal 2008; Okamoto \etal 2008).

In our earlier work (TSC06) we incorporated the merger and feedback
AGN model of  SO04 into a large cosmological smooth particle
hydrodynamic simulation.  However, our analysis of the clustering
properties of optically-selected AGNs was constrained to a single
luminosity bin and did not consider redshift evolution in any
significant depth.  In view of several recent ground-based surveys, it
is therefore timely to reanalyze our simulation to address both the 
luminosity
dependence and redshift evolution of clustering.  Furthermore, the
advent of new X-ray selected AGN catalogs with optical follow-up also
allows us to present an analysis of the clustering of an X-ray selected catalog.

We stress that the aim of this paper is not to encourage support for one
quasar model over another, but rather to examine whether the details of
the faint part of the light curve are actually needed to accurately
predict currently observed clustering statistics. Since the SO04 model
does not include a low luminosity accretion period we can indirectly
constrain the importance of this epoch to quasar clustering behavior.
This should not be interpreted as constraining whether such a period
does actually occur, or for that matter, the relative length of such a
period.

The layout of the paper is as follows: in \S 2 we give a summary 
of our simulations and  overall method. In \S 3 we present a detailed
analysis of clustering at $z=1.5,$ 2.0, and 3.0 and
the dependence of this clustering on luminosity. 
In \S 4 we examine the clustering properties of X-ray selected
AGN, again as a function of redshift and luminosity.  We
close with  a brief discussion in \S 5. Throughout  the  paper we
consider a pre-WMAP3 (Spergel \etal 2003) $\Lambda$CDM model with
parameters $h=0.7$, $\Omega_0$ = 0.3, $\Omega_\Lambda$ = 0.7,
$\Omega_b = 0.046$, $\sigma_8 = 0.9$, and $n=1$, where $h$ is the
Hubble constant in units of 100 km s$^{-1}$ Mpc$^{-1}$, $\Omega_0$,
$\Omega_\Lambda$, and $\Omega_b$ are the total matter, vacuum, and
baryonic densities in units of the critical density, $\sigma_8^2$ is
the variance of linear fluctuations on the $8 h^{-1}{\rm Mpc}$ scale,
and $n$ is the ``tilt'' of the primordial power spectrum. While we 
consider only a fixed $\sigma_8$ in our simulation, the overall 
impact of changing $\sigma_8$ on fitted correlation functions 
($\xi(r)=(r_0/r)^\gamma)$ is to change the correlation length $r_0$. For 
correlation functions with $\gamma\approx2$, increasing $\sigma_8$ by a 
factor of $f$ will increase the correlation length of unbiased tracers 
of mass by the same factor. Throughout the paper the
Eisenstein \& Hu (1999) transfer function is used and we quote 
all distances in comoving coordinates.

\section{Simulation Method and Quasar Modeling}

We consider two simulations in this study. The first is a ``fiducial" run
containing star formation and our model of AGN outflows (TSC06) in a
periodic cube 146 $h^{-1}$ Mpc on a side, containing $2\times
640^3$ particles. With these choices the dark-matter particle mass is $1.9
\times 10^{8} \msun$ and the gas particle mass is $2.7 \times 10^{7}
\msun.$ The second simulation, which we call the ``comparison" run, uses
$2\times320^3$ particles in a periodic cube of size 73 $h^{-1}$ 
Mpc, which matches the
particle mass in the fiducial run, and includes star formation but not AGN
outflows (Scannapieco \etal 2008).  Both simulations were conducted with a
parallel OpenMP based implementation of the ``HYDRA'' code (Thacker \&
Couchman 2006)  that uses the Adaptive Particle-Particle, Particle-Mesh
algorithm to calculate gravitational forces (Couchman 1991), and the
smooth particle hydrodynamic method to calculate gas forces (Lucy 1977;
Gingold \& Monaghan 1977).  Due to computational cost as well as the
limitations of our modeling, both simulations were halted at $z=1.2$.

Our method, as outlined in TSC06, associates quasar-phase AGN with
galaxy mergers, which are  tracked within the simulations
by identifying gas groups and applying  group number labels to their
particles. Mergers are groups for which at least 30\% of the
accreted mass  does not come from a single massive progenitor, and
we calculate the mass of the
associated black hole, $M_{\rm  BH},$ using the circular velocity of
the new system, $v_{\rm c}$, and   the observed $M_{\rm BH}- v_{\rm
c}$ relation (Merrit \& Ferrarese 2001; Tremaine \etal 2002; Ferrarese
2002). We note that observational evidence of the universality of this 
relationship is still restricted to low redshift systems, and there is 
modest evidence that the normalization changes somewhat with redshift (\eg 
Woo \etal 2008). Continuing, our modeling approach  yields   \be   
M_{\rm BH} = 2.8 \times 10^8
\left( \frac{v_{\rm c}}{300 \, {\rm km} \,{\rm s^{-1}}} \right)^5, \ee
where $v_{\rm c}$ is estimated from    \be v_{\rm c} =
\left[\frac{4\pi}{3} G \rho_{\rm v}(z) r_{\rm v}^2 \right]^{1/2},
\ee and $G$ is the gravitational constant, $\rho_{\rm v}(z)$ is the
virial density as a function of redshift, and $r_{\rm v}$ is the
implied virial radius for a group of $N$ gas particles with mass
$m_{\rm g}$ \be   r_{\rm v}=\left[ {N  m_{\rm g} \Omega_0/\Omega_b
\over 4/3 \pi \rho_{\rm v}(z)}  \right]^{1/3}.    \ee

In keeping with the model outlined in Wyithe \& Loeb (2002) and SO04,
we   assume that for each merger the accreting black hole shines at
its  Eddington luminosity ($1.2 \times 10^{38}$ ergs s$^{-1}$
$\msun^{-1}$) for a time taken to be a fixed fraction, $0.055,$ of the
dynamical time of the system, $t_{\rm AGN} = 0.055 r_{\rm v}/v_{\rm c}
= 5.8 \times 10^{-3} \Omega(z)^{-1/2} H(z)^{-1}$. We have shown in
earlier work (TSC06) that apart from a small discrepancy at the most
luminous end of the luminosity function, these simple assumptions lead
to  a model that reproduces the observed AGN luminosity function as
well as the predictions of SO04.  Furthermore, we were able to
demonstrate that  this discrepancy can be explained in terms  of the
relative efficiency of shock heating on substructure, and corrected
for if necessary by post-processing our simulations, as discussed in
further detail below.  An initial  analysis of clustering properties
was also in close agreement with  observations, particularly the
small-scale, $r \lesssim 1 h^{-1}$ Mpc, clustering of quasars (\eg  Hennawi 
\etal 2006).

Perhaps the most problematic aspect of this model is that it makes no 
distinction between AGN formed by gas-rich ``wet'' mergers versus those 
formed by gas-poor ``dry'' mergers. As we implement star formation on 
the basis of a simple merger model and ignore the quiescent mode of star 
formation, it is difficult for us to make this distinction with any 
confidence. However, while there is significant evidence that at low 
redshifts the quiescent mode of star formation dominates (Noeske \etal 
2007), at higher redshifts there is good reason to believe that mergers 
are necessary to fuel observed high star formation rates (\eg Erb 
2008). We 
can also appeal to the fact that while at $z=0$ dry mergers are 
important, they will be less so at $z=1.2.$ For example, the 
semi-analytic estimates presented in Hopkins \etal (2007), in particular 
their Figure 5, show that that the ratio of gas-rich to gas poor $4 
\times 10^{12} M_\odot$ mergers, ranges from 10:1 at $z=2$, to 2:1 at 
$z=1$, indicating that gas-poor mergers are relatively unimportant before
our final redshift.

\begin{figure*}
\epsscale{0.9}
\plotone{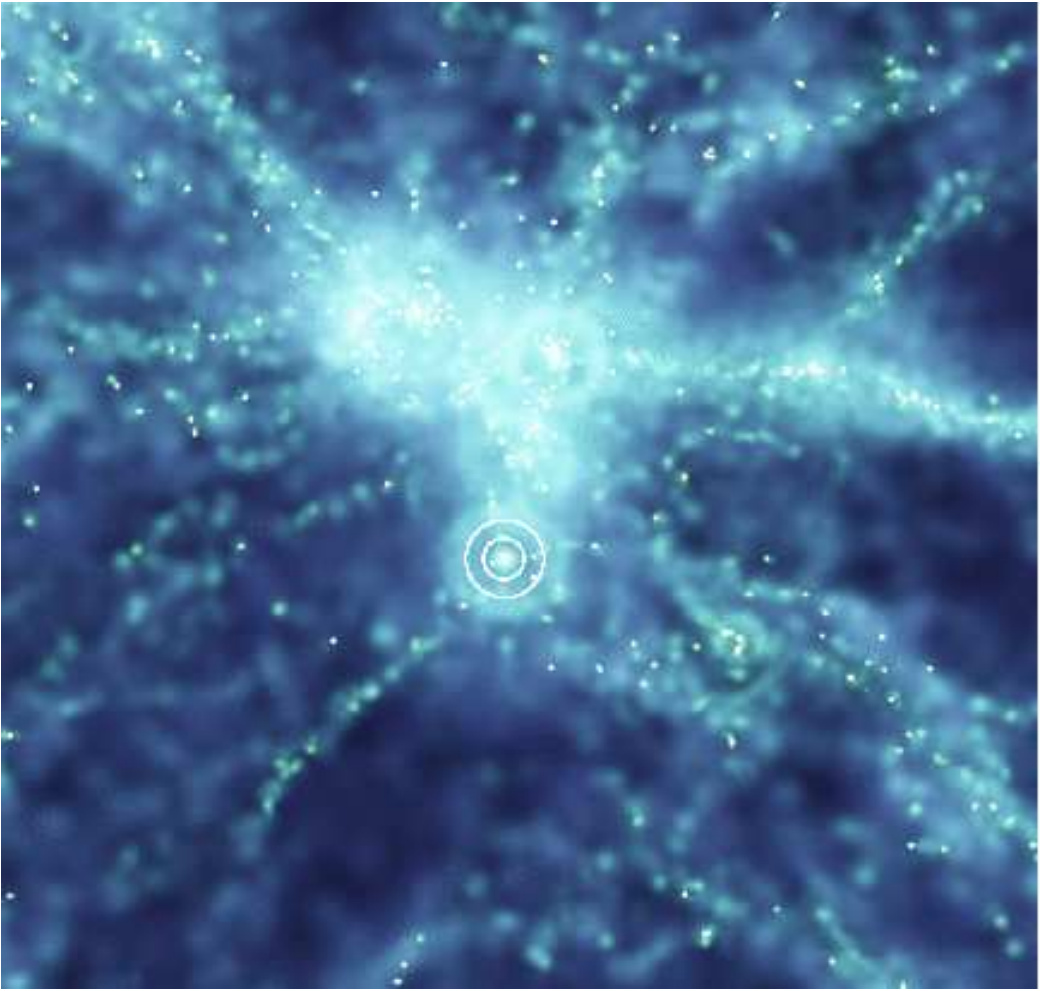}
\caption{Schematic representation of the impact of outflows on the local 
density of gas. The region shown is 12 $h^{-1}$ Mpc across (comoving), 
at a redshift of z=1.59. The the virial radius, $r_v$, of a system with 
a baryonic mass of $6\times10^{10}\, M_\odot$ is represented by the 
inner circle, the outer circle corresponds to $2r_v$, denoting the 
launching point of an outflow for this system. A number of outflow 
events with different characteristic radii are visible within this small 
volume.
}
\label{fig:scheme}
\end{figure*}

In the calculation of wind velocity we assume that a fixed fraction
$\epsilon_{\rm k} =0.05$ of the bolometric energy of each AGN is
channeled into a kinetic outflow, while the remainder is emitted as
light. While there is much debate about variability of this  value on
a system-by-system basis, our choice is consistent with other
literature estimates  (\eg Furlanetto \& Loeb 2001; Nath \&
Roychowdhury 2002), as well as observations (Chartas \etal 2007).  If
we restrict ourselves to the consideration of systems with large
bulges, then the resulting level of kinetic energy input is
considerably  greater than  that from supernovae and stellar winds
(\eg Kravtsov \& Yepes 2000; Tozzi \etal 2000; Brighenti \& Mathews
2001;  Babul \etal 2002; Tornatore \etal 2004).  Using the Eddington
luminosity, associated dynamical time and the wind  efficiency, each
AGN outflow is launched with a wind energy of   \be   E_{\rm k}=6
\times10^{36}\; \left({ M_{\rm bh} \over  \msol}\right) \left({t_{\rm
d} \over  {\rm s}}\right) \;{\rm ergs}.
\label{eq:ekin}
\ee   Since there are considerable uncertainties about the precise
geometry of  AGN outflows, we have chosen to use a spherical shell to
represent the  outflow. Even strongly bipolar systems will tend to
release an  ellipsoidal cocoon of gas (\eg Begelman \& Cioffi 
1989; Yamada \etal 1999), so this 
approximation is reasonable.  We thus model each expanding outflow as
a spherical shell at a radius $2r_{\rm v}$  which is created by
rearranging the gas between $r_{\rm v}$ and $2r_{\rm v}$ which lies
below a density threshold of $2.5\rho_{\rm v}.$ In Figure 
\ref{fig:scheme} we present a 
plot of local gas density in a 12 
$h^{-1}$ Mpc region with the position of the virial and launching radii 
indicated.
The radial velocity of
the shell $v_s$ is set by ensuring that the sum of the thermal and
kinetic wind energies is equal to $E_{\rm k}-E_{\rm grav}$ where
$E_{\rm grav}$  is the potential energy change required to move the
particles to  $2r_{\rm v}$. The post-shock temperature of the wind is
given by   
\be 
 T_s = 13.6\, {\rm K} \left( { v_s \over {\rm km}\,
{\rm s}^{-1}} \right)^2.    
\ee    
This model produces a
level of preheating in galaxy clusters  and groups that is in good
agreement with observations as discussed in TSC06, to which the
reader is referred for further details about our simulations.

\section{Optically-Selected AGN}

\subsection{Optical Quasar Luminosity Function Revisited}

\begin{figure*}[t]
\plotone{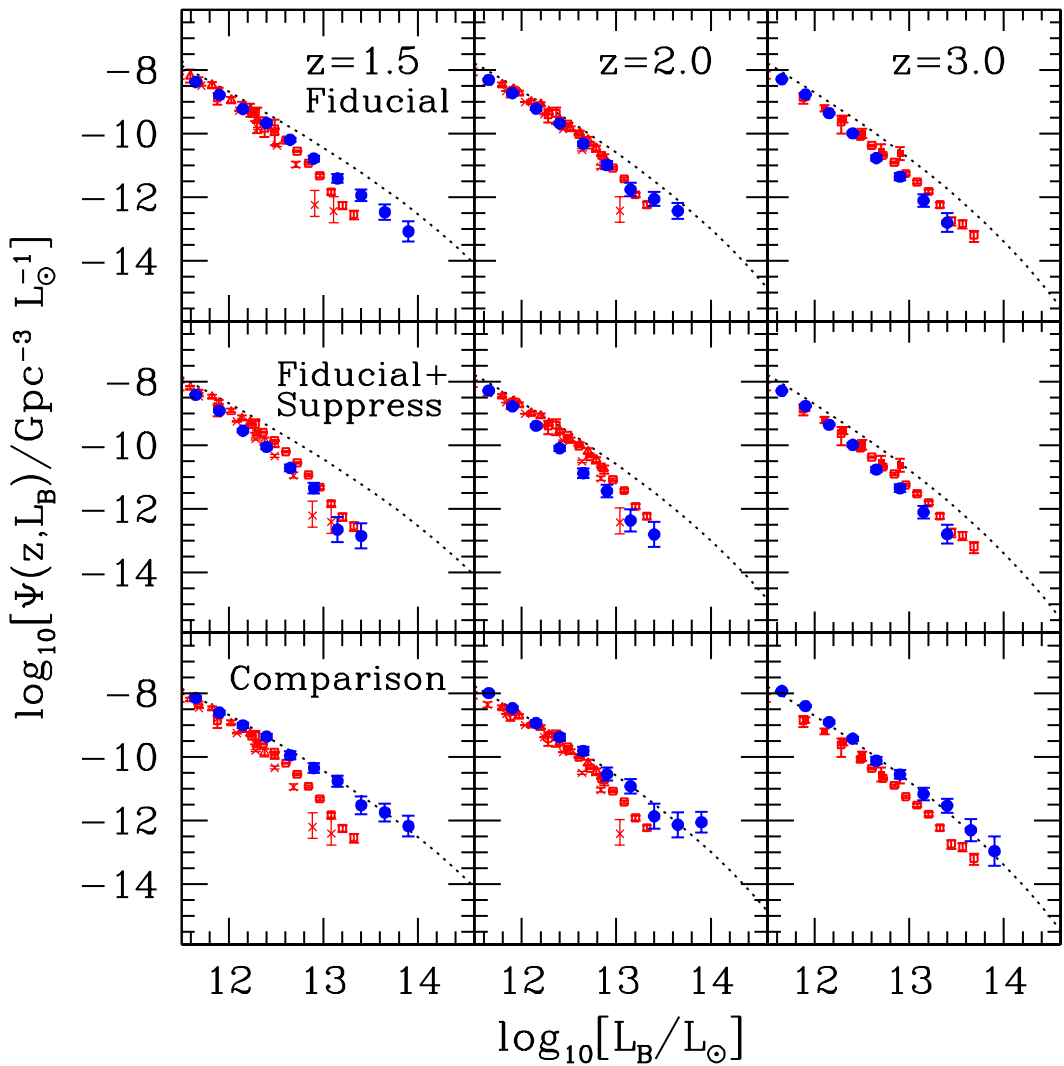}
\caption{Evolution of the B-band quasar luminosity function. The 
simulation results are given by the solid circles, while the dotted
line is the simple estimate from the  analytic model of Wyithe \& Loeb
(2003).    From left to right the columns give results at redshifts of
$1.2-1.75,$  $1.75-2.25,$  and $2.25-4.0$.  From top to bottom, the rows
show results from the fiducial run, the fiducial run with additional
suppression imposed (by removing neighboring systems below a heating 
threshold), and  the comparison run.  In all panels error bars are 1-sigma Poisson estimates.
The observational data are taken from Croom \etal (2004, crosses) Richards \etal 
(2005, open triangles), Richards \etal (2006, open squares), Wolf \etal (2003, open circles),
and Siana \etal (2008, filled squares).}
\label{fig:lum}
\end{figure*}

As a test of our overall approach, our first step is to repeat the
optical luminosity function analysis of TSC06,  but for both
simulations and for ranges of redshifts binned so as to make them most
useful for comparisons with clustering measurements.  As
previously, we construct the luminosity function by binning in
luminosity and redshift.  We calculate the number of quasars in each
bin times the total time these objects are shining, and divide by the
time interval, the width of the bin, and the volume of the simulation.
That is for a given redshift bin $i$ and a given  luminosity bin $j$
the luminosity function is simply   
\be   \Psi_{i,j} = \frac{1}{V
\Delta t_i \Delta L_{B,j}} \sum_{k  \in {\rm bin_{i,j}}} t_{{\rm AGN},k}, 
\ee
where the sum is over the lifetimes of
all quasars with redshifts and luminosities
associated with the $i$, $j$ bin, which spans  a time interval $\Delta
t_i$ and a range of luminosities  $\Delta L_{B,j}.$ The resulting
luminosity functions for the fiducial AGN-feedback run and the comparison
run are shown in Figure \ref{fig:lum}, in which the error bars are 1-sigma
estimates,  computed
as $\Delta \Psi_{i,j} = \Psi_{i,j} [1 \pm (1 + N_{i,j})^{-1/2}],$ where $N_{i,j}$
is the number of quasars contributing to bin $i$,$j$.
As discussed in TSCO6, our
fiducial model shows a clear turn-down in the number of $L_B \ge
10^{13}$ quasars at $z < 2$, which parallels the observational trend,
but still overestimates the number of luminous and
low-redshift quasars due to numerical effects.  Likewise, the
luminosity function in the no-feedback simulation continues to rise at
low redshift, increasing along with the halo merger rate as discussed
in Wyithe \& Loeb (2003).  Thus, this simple model fails to reproduce
the drop in the number density of $z \leq 2$  quasars as discussed in
SO04 and Scannapieco \etal (2008). 
 
Finally, we include a luminosity function calculated to match the
shock behavior in the semi-analytic SO04 model. This was achieved by
removing neighboring objects from the simulation that are found inside
a shock radius calculated using the SO04 model (see TSC06 for an
extended discussion of this analysis).  After we apply this algorithm,
the simulation results much more closely match  the observations.
  
\subsection{Dependence of the Correlation Function on Redshift and Luminosity}

Having outlined the successes and limitations of our model in
reproducing the observed number density of  quasars, we next move on
to a detailed study of their spatial distribution.  Here our primary
tool is the real-space auto-correlation function,
calculated as   
\be   
\xi_{\rm qq}(r,z,L) + 1 = {DD(r,z,L) \over RR(r,z,L)},   
\ee   
where $DD(r,z,L)$ is the number of pairs at a
given comoving distance within a  given redshift bin and with a
luminosity within a given interval, and $RR(r,z,L)$ is the average
number of such pairs that would be found at  this separation in a
random distribution.   Here we have correlated
all quasars within each redshift bin,
regardless of  whether the two objects are shining simultaneously.
While this vastly improves the statistical signal, the use of a
relatively large redshift window places a lower limit 
on the spatial scales that we can study, because  the peculiar motions 
can shift the positions of the quasars during
the finite time window associated with each bin.    For our choices
of redshift intervals, and estimating typical peculiar velocities of
quasars at $\approx 300$ km/s, this places a lower limit of $0.5$
 $h^{-1}$ Mpc.  Note that intrinsic velocity dispersion is estimated 
from the properties of the halos in which the majority of 
quasars are contained, and is somewhat smaller than observed pairwise
dispersions (\eg da Angela \etal 2005), which include both intrinsic
and observational errors.

The finite volume of our simulation
places an upper limit on the distance we can study of approximately 1/5
the box size (Scoccimarro 1998;  Szapudi \etal 1999), which
corresponds to  30  $h^{-1}$ Mpc in the  AGN feedback run and
15  $h^{-1}$ Mpc in the comparison run. To be especially
conservative we use a  cutoff radius  of 10$h^{-1}$ Mpc for most of
our results, which allows a direct  comparison of  the fiducial and
comparison runs. For the fiducial run alone we also  examine the
impact of changing to a   25$h^{-1}$ Mpc cutoff.
In each bin the error bars have be computed using a simple 1-sigma
Poisson estimate of 
$\Delta \xi_{\rm qq} (r,z,L) = \xi_{\rm qq} (r,z,L)[1 \pm (1 +DD(r,z,L))^{-1/2}]$

\begin{figure*}[t]
\plotone{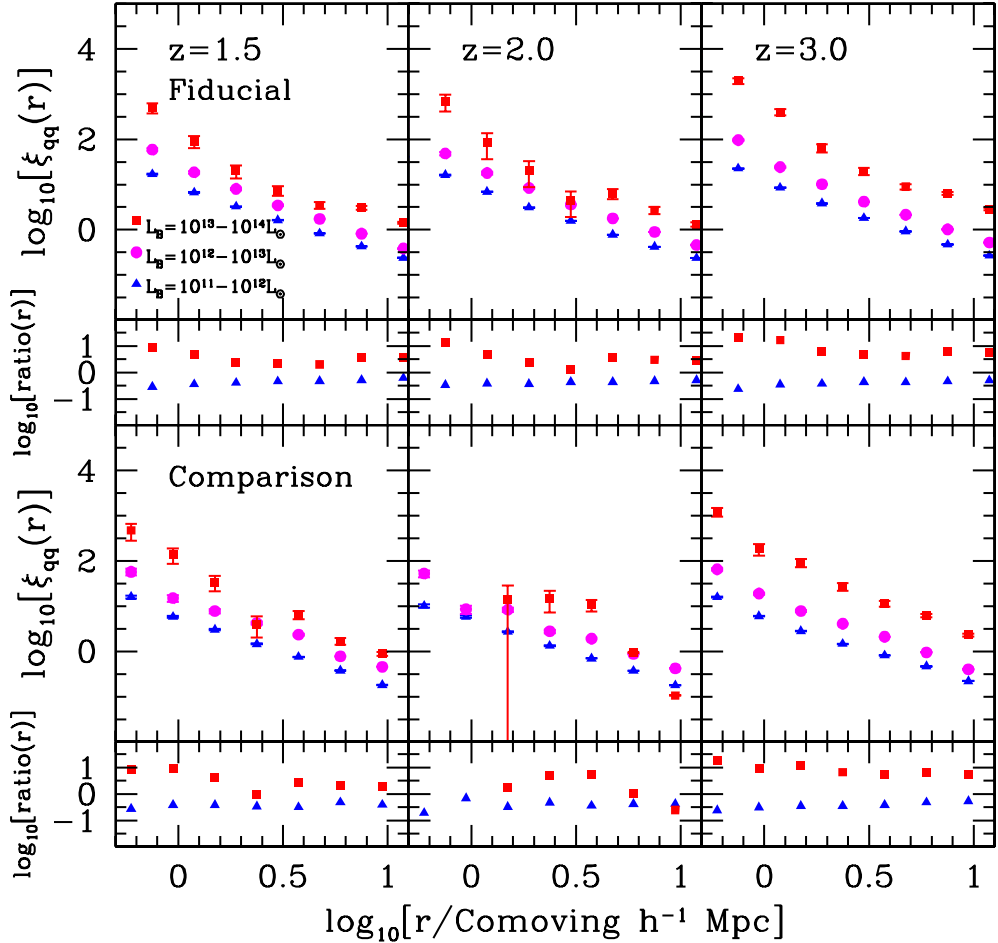}
\caption{Luminosity and redshift dependence of  the quasar
auto-correlation function, $\xi_{\rm qq}(r)$.  The top row shows the
results our AGN-feedback simulations, as calculated by partitioning
quasars into bins with  $L_B = 10^{11}-10^{12} L_{\odot,B}, $ $L_B =
10^{12}-10^{13} L_{\odot,B},$  and $L_B = 10^{13}- 10^{14}
L_{\odot,B}.$  In the second row we show  the ratio of $\xi_{\rm
qq}(r)$ for $L_B = 10^{13}- 10^{14} L_{\odot}$ quasars over
$\xi_{\rm qq}(r)$ for $L_B = 10^{12}- 10^{13} L_{\odot}$ quasar
(squares) and the ratio for $L_B = 10^{11}- 10^{12} L_{\odot}$
quasars over $L_B = 10^{12}- 10^{13} L_{\odot},$ quasars again from
the AGN-feedback simulation.   In the third row we show $\xi_{\rm
qq}(r)$ from our  no feedback comparison simulation, with symbols as
above, and the ratios of the correlation functions from this run are
given the bottom row.  As in Figure 2, from left to right each column
shows the results from $z=1.2-1.75,$  $z=1.75-2.25,$  and $z=2.25-4.0,$
and all error bars are 1-sigma Poisson estimates.}
\label{fig:xi}
\end{figure*} 

With these limitations in mind, in Figure \ref{fig:xi} we plot the
correlation function of simulated quasars, dividing our sample into
three luminosity bins from  $L_B = 10^{11}-10^{12} L_{\odot,B}, $ $L_B
= 10^{12}-10^{13} L_{\odot,B},$  and $L_B = 10^{13}- 10^{14}
L_{\odot,B}.$    Focusing first on the AGN feedback run, the most
striking feature of this plot is the relative lack of clustering
at large separations and low redshifts, which occurs even
though our model assumes accretion at the Eddington rate for all
active black holes.   Perhaps contrary to initial expectations, this
weak dependence blankets the range of redshifts and separations that
are best constrained observationally, suggesting that complex
accretion histories may not play a key role in explaining current
optical measurements.

Note, however, that very different behavior occurs both at the
smallest separations and  at the highest redshifts, with $\xi_{\rm
qq}$ showing a strong luminosity dependence in both these regimes.
Each of these enhancements is likely to be caused by different
processes.    At small separations, the strong dependence is likely to
be a manifestation of so-called ``one-halo effects''  (\eg Berlind \&
Weinberg 2002; Bullock \etal 2002;  van den Bosch \etal 2003;
Magliocchetti \& Porciani 2003) which take place when
gravitationally-bound objects orbit each other within the same
potential,  adding significantly to the correlation function at
distances smaller than the virial radius.   This small-scale upturn,
which  has been confirmed observationally for SDSS quasars (Hennawi
\etal 2006; Serber \etal 2006; Myers \etal 2007b), occurs at the
radius corresponding to the maximum apocenter of such
gravitationally-bound pairs, which in turn corresponds  to the virial
radius of the halos in which they are contained.   As the most
luminous AGN  live in the deepest  gravitational potential wells in
our model,  this means that small-scale clustering is naturally
enhanced for these objects, causing a break in $\xi_{\rm qq}(r)$ that
occurs at larger radii for more luminous objects.  

On the other hand,
the luminosity dependence seen at $\gtrsim 2$ Mpc
$h^{-1}$ in the $z=3$ bin is on such large scales that it can not be
due to this effect.  Instead this enhancement is likely to be a result
of ``geometrical bias,'' which is caused by the statistics of peaks
within a Gaussian random field (\eg Kaiser 1984; Bardeen \etal 1986;
Mo \& White 1996; Porciani \etal 1998). Note that on these scales the
increase in $\xi_{\rm qq}(r)$ is independent of distance, further
pointing to this origin.

Similar trends are apparent in the comparison run. Again there is
little  luminosity dependence at $z \leq 2$ and $r \geq 1$ 
$h^{-1}$ Mpc. Also as in the  feedback
case, strong luminosity dependence is detected in the two regimes that
are least constrained observationally: small-scale $\leq 2$ Mpc
$h^{-1}$ clustering, which is likely to be dominated by one-halo
effects, and larger-scale high-redshift clustering,  which is likely
to be dominated by geometric bias.

\begin{figure*}[t]
\plotone{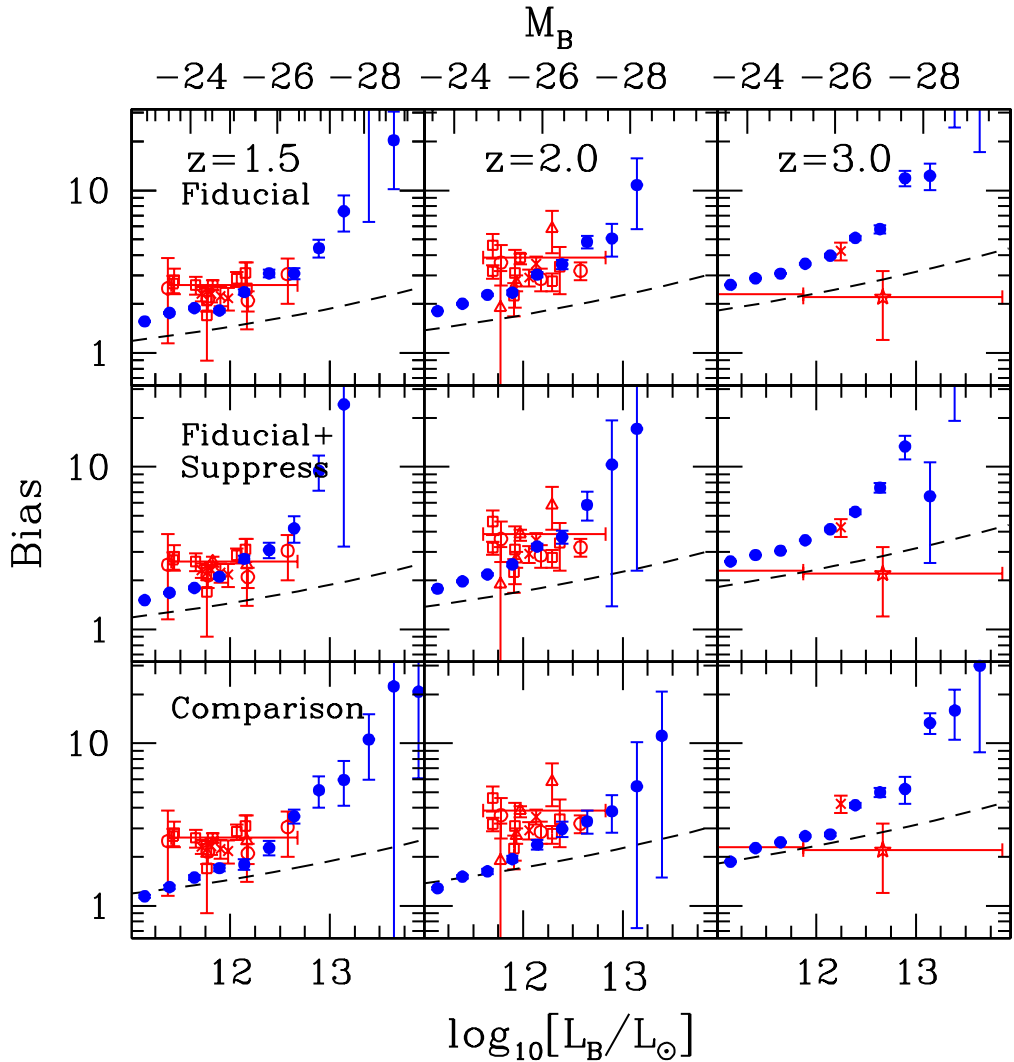}
\caption{{\em Top:} Bias of quasars as a function of redshift and
B-band luminosity.  As in Figure 2, simulation results are given by
the solid points with 1-sigma Poisson error bars,  and the  dashed line is
 the simple Sheth \etal (2001)
estimate of the bias as described in the text.
The open squares corresponds to the  observational data points, which
are taken from  Porciani \etal (2004) and Porciani \& Norberg (2006),
open triangles; Croom \etal (2005), crosses; Adelberger \& Steidel
(2005), stars; and Myers \etal (2006; 2007a), squares.  From top to
bottom, the rows correspond to the fiducial run, the fiducial run with
additional suppression, and the comparison run. Columns correspond to
redshift bins as in Figures 2 and 3.}
\label{fig:bias}
\end{figure*}

To quantify our results further we computed the bias of quasars as
function of $L$ and $z$ as  
\be 
b^2(L,z) =  \frac{\sum_k w(r_k,L,z)
\xi_{\rm qq}(r_k,L,z) \xi_{\rm DM}(r_k,z)^{-1}}  {\sum_k w(r_k,L,z)},
\ee 
where $\xi_{\rm qq}(r_k,L,z)$ is the quasar auto-correlation
function in  a radial bin $k$ as a function of luminosity and redshift,
$\xi_{\rm DM}(r,z)$ is the linear dark matter correlation function
extrapolated to
the redshift bin of interest, $w(r_k,L,z)$ is a weighting function
that counts the number of pairs  contributing to each value of
$\xi_{\rm qq}(r_k,L,z)$, and we average over the interval from $r=$
1.0 to 10  $h^{-1}$ Mpc.  Error bars are again computed from
Poisson estimates.  A selected list of the
computed bias values is given in Table \ref{table1} and the full 
data-set 
is plotted in
Figure \ref{fig:bias}, in which we have also compiled  results from
current surveys, extrapolating to the B-band with a spectral  slope of
$\alpha_\nu = -0.5,$ (Wyithe \& Loeb 2005; TSC06).  Note however that
these surveys do not necessarily estimate bias at exactly the same
range of separations as we have used. In particular, this plot
includes points from  Porciani \etal (2004) and Porciani \& Norberg
(2006), measured at $\approx 2-20$ comoving $h^{-1}$ Mpc,  Adelberger
\& Steidel (2005), measured at $\leq 30$ comoving $h^{-1}$ Mpc,  Croom
\etal (2005), measured at $\approx 1-20$ comoving $h^{-1}$ Mpc, and
Myers \etal (2006; 2007a),  measured from $\approx 1-100$ $h^{-1}$ Mpc. 
Finally, for comparison purposes, we have also added a
simple analytic estimate of the bias expected in the no feedback case,
in which black hole mass can be directly related to the halo velocity
dispersion and hence to the halo mass as in Wyithe \& Loeb (2002).  In
this case, 
\be 
 b(L,z) = 1 + \delta_{0,c}^{-1} \left[ \nu'^2 + b
\nu'^{2(1-c)} - \frac{\nu'^{2c}/\sqrt{a}}{\nu'^{2c}+b(1-c)(1-c/2)}
\right],
\label{eq:SMTbias}
\ee where $a=0.707$, $b=0.5$, $c=0.6$ , $\delta_{0,c}= 1.69,$
$\nu'=a^{1/2}\delta_{0,c} D(z)^{-1} \sigma^{-1}(M_{\rm halo})$, $D(z)$ is
the linear  growth factor, $\sigma(M_{\rm halo})$ is the $z=0$
variance on the halo mass  scale, $M_{\rm halo},$ corresponding to a
given quasar luminosity (Sheth \etal  2001;  see also Mo \& White
1996; Jing 1999; Scannapieco \& Barkana 2002), in the case in which
gas accretion and dark matter collapse occur simultaneously, maintaining
the cosmological ratio at all times.

\begin{table*}[t]
\begin{center}
\caption{Selected correlation lengths and biases for the fiducial and
comparison  runs as a function of redshift, luminosity, and
selection band. Optically-selected values are given first. A key of `F'
corresponds to the  fiducial run, while `C' to the comparison
run. Bolometric luminosities  associated with each bin are given,
along with the associated B-band  luminosity (for the optical
catalog) and hard X-ray luminosity (for the  X-ray
catalog). Correlation lengths are quoted to 2 significant figures
without errors since statistical errors will be significantly smaller
than  systematic errors from the binning procedure.}
\fontsize{10}{10pt}\selectfont
\begin{tabular}{lcccccccc}
\hline\hline
Run  & $z$ &log($L_{Bol}$/$L_\odot)$ &
log($L_B$/$L_\odot$) &  $M_B$  & $L_X$ & $r_{cut}$ & $r_0$ &
b  \\
 & & &  & & (erg s${}^{-1}$) & ($h^{-1}$ Mpc) & ($h^{-1}$
Mpc) & \\
 \hline \vspace{1mm}
F & 1.5 & 12.7 & 11.6 & -24.3 & & 10 & 5.2 & 1.9 \\
F & 1.5 & 12.7 & 11.6 & -24.3 & & 25 & 4.4 & 1.6 \\
C & 1.5 & 12.7 & 11.6 & -24.3 & & 10 & 4.1 & 1.5 \\
F & 2.0 & 12.7 & 11.6 & -24.3 & & 10 & 5.3 & 2.3 \\
F & 2.0 & 12.7 & 11.6 & -24.3 & & 25 & 4.6 & 2.0 \\
C & 2.0 & 12.7 & 11.6 & -24.3 & & 10 & 3.8 & 1.6 \\
F & 3.0 & 12.7 & 11.6 & -24.3 & & 10 & 5.4 & 3.1 \\
F & 3.0 & 12.7 & 11.6 & -24.3 & & 25 & 5.3 & 3.0 \\
C & 3.0 & 12.7 & 11.6 & -24.3 & & 10 & 4.3  & 2.5 \\
F & 1.5 & 13.2 & 12.1 & -25.6 & & 10 & 6.4  & 2.4 \\
F & 1.5 & 13.2 & 12.1 & -25.6 & & 25 & 6.3 & 2.2 \\
C & 1.5 & 13.2 & 12.1 & -25.6 & & 10 & 4.8  & 1.8 \\
F & 2.0 & 13.2 & 12.1 & -25.6 & & 10 & 7.2  & 3.0 \\
F & 2.0 & 13.2 & 12.1 & -25.6 & & 25 & 6.2  & 2.6 \\
C & 2.0 & 13.2 & 12.1 & -25.6 & & 10 & 5.6  & 2.4 \\
F & 3.0 & 13.2 & 12.1 & -25.6 & & 10 & 7.1  & 4.0 \\
F & 3.0 & 13.2 & 12.1 & -25.6 & & 25 & 7.4  & 4.0 \\
C & 3.0 & 13.2 & 12.1 & -25.6 & & 10 & 4.6  & 2.7 \\
F & 1.5 & 13.4 & 12.4 & -26.2 & & 10 & 7.8  & 3.1 \\
F & 1.5 & 13.4 & 12.4 & -26.2 & & 25 & 5.7  & 2.1 \\
C & 1.5 & 13.4 & 12.4 & -26.2 & & 10 & 6.3 & 2.3 \\
F & 2.0 & 13.4 & 12.4 & -26.2 & & 10 & 8.6  & 3.5 \\
F & 2.0 & 13.4 & 12.4 & -26.2 & & 25 & 6.9 & 2.9 \\
C & 2.0 & 13.4 & 12.4 & -26.2 & & 10 & 6.7 & 3.0 \\
F & 3.0 & 13.4 & 12.4 & -26.2 & & 10 & 9.1 & 5.1 \\
F & 3.0 & 13.4 & 12.4 & -26.2 & & 25 & 8.3 & 4.4 \\
C & 3.0 & 13.4 & 12.4 & -26.2 & & 10 & 7.5  & 4.1 \\
F & 1.75 & 10.0 &  & & $3.2 \times 10^{42}$ &  10 & 3.9  & 1.6 \\
C & 1.75 & 10.0 &  & & $3.2 \times 10^{42}$ &  10 & 3.1  & 1.2 \\
F & 1.75 & 11.3 &  & & $3.2 \times 10^{43}$ &  10 & 3.6  & 1.5 \\
C & 1.75 & 11.3 &  & & $3.2 \times 10^{43}$ &  10 & 3.2   & 1.3 \\
F & 3.0  & 11.3 &  & & $3.2 \times 10^{43}$ &  10 & 3.3  & 2.0 \\
C & 3.0  & 11.3 &  & & $3.2 \times 10^{43}$ &  10 & 3.2  & 1.8 \\
F & 1.75 & 12.6 &  & & $3.2 \times 10^{44}$ &  10 & 5.2 & 2.1 \\
C & 1.75 & 12.6 &  & & $3.2 \times 10^{44}$ &  10 & 4.5 & 1.7 \\
F & 3.0  & 12.6 &  & & $3.2 \times 10^{44}$ &  10 & 5.5 & 3.1 \\
C & 3.0  & 12.6 &  & & $3.2 \times 10^{44}$ &  10 & 4.5 & 2.5 \\
\hline
\smallskip
\label{table1}
\end{tabular}
\end{center}
\fontsize{11}{11pt}\selectfont
\end{table*}

Expressing our $\gtrsim 1$ $h^{-1}$ Mpc  results as a bias allows for
easy quantification of the trends seen in Figure \ref{fig:xi},  as
well as comparisons with observations. Focusing first on the AGN
feedback  results, we find that bias increases by no more than $\approx 50 \%$
over the range of luminosities and redshifts probed by current
surveys. The observational data indicate that current clustering bias
measurements (Porcani \etal 2004; Adelberger \& Steidel 2005; Croom
\etal 2005; Porcani \& Norberg 2006; Myers \etal 2006, 2007a) do not
provide any significant statistical constraints above  $\log_{10}
(L_B/L_\odot)=12.5$. Indeed, a sample large enough to detect
luminosity dependence of bias with $\Delta b\simeq 1$ at  a $3\sigma$
confidence level, given current detection limits (such as 2QZ), would
require an all-sky measurement (Porcani \& Norberg 2006). At
$z=3$, there is a suggestion in our simulations
that bias changes significantly with luminosity above
$\log(L_B)={12}$, but this regime is poorly-constrained
observationally.

While this mismatch between the observed range of redshifts and
luminosities and the regime in which we expect strong luminosity
dependence appears initially to be somewhat of a conspiracy, it can be
understood naturally as a consequence of our feedback modeling.  As
discussed in detail in Scannapieco \etal (2005) and TSC06, AGN
outflows act to impose a maximum halo mass, above which gas is unable
to cool efficiently, suppressing further generations of galaxies and
quasars.   Furthermore, as radiative cooling is
proportional  to the square of the gas density, cooling is much more
efficient in dense,  high-redshift structures than it is at lower
redshifts.   This means that the ``quenching threshold'' \ie the mass at 
which AGN is shut down (Faber \etal
2007) should decrease with time, with the strongest AGN quenching
galaxy formation even at high redshifts, but smaller smaller  AGN
quenching galaxy formation only at low redshifts.  At the same
time  the hierarchical nature of dark-matter driven gravitational
collapse means that the nonlinear mass scale {\em increases} with
time, as ever-larger structures  collapse and virialize.


This means that AGN are naturally divided into two regimes.  At high
redshift,  the characteristic luminosity of active black holes
brightens along with the nonlinear mass scale, while at low redshift,
their characteristic luminosity fades along with the quenching
threshold.  The $z \approx 2$ peak of AGN activity then marks a
distinct transition between hierarchical and anti-hierarchical
formation, which occurs when the quenching threshold drops below the
nonlinear mass scale.   Thus the majority of AGN formed at
redshifts below the peak of AGN activity, meaning those that are
easiest to  observe and quantify, are naturally found in halos with
masses well below the nonlinear mass scale. As can been seen from eq.\
(\ref{eq:SMTbias}), these low masses are very weakly biased,
as $b^2$ is a strong function of $\nu'$ when $\nu' \gtrsim 1$, but
almost a constant when $\nu' \lesssim 1.$

Another important feature of our feedback model is that it produces
halo biases  that are systematically offset from the no-feedback case.  
This is because feedback acts to slow accretion even
before the quenching threshold is passed, meaning that  each halo
hosts a somewhat less massive black hole  than it would have in the
absence of feedback. Thus, for a fixed luminosity, each AGN is shifted
to a somewhat more massive, and hence more clustered, dark matter
halo, and the typical increase in bias over the no-feedback case is 
about 30\%. At the faint end this corresponds to an increase over the 
analytic estimate of about a factor of $\approx 2$.
The quasars points from our simulation are also offset
from the simple Sheth \etal (2005) estimates, again because gas
accretion lags behind the dark matter collapse in the simulation.

It is important to note that  post-processing our results to correct
for  the handful of very luminous, low-redshift AGN that result from
inefficiency in shock heating in the simulation has very little effect
on any of these trends, despite it leading to an almost perfect match
of the  luminosity function.  As shown in the center row, removing
these objects only impacts the $L_B \gtrsim 10^{13} L_\odot$
measurements in the lowest redshift bins, primarily increasing the
already large error bars by further lowering the number density of
these objects.

In the comparison run, on the other hand, the overall bias at each
mass scale is somewhat lower than in either of the other two
cases. Again this is because while gas accretion and cooling still
take time, this time is much less than the $2-5$ Gyr Hubble times at
these epochs, meaning that the  relationship between black hole mass
and halo mass is more in line with that expected purely  from the dark
matter distribution.  Note that even in this run however, there is
very little evolution in clustering over the {\em observed} range of
luminosities.   This is because even though no feedback is included,
leading to a fair number of large and biased low-redshift AGN in the
simulation, the lack of  low redshift quasars in the data in the
observations means these objects simply do not exist in nature, and
thus can not be compared to our predictions.  However, even in the
absence of feedback, the significant cooling time associated with
large objects means that gas accretion trails dark matter collapse.
Thus means that quasars are found in higher-mass halos, and hence are
more clustered, than one would expect   in a simple  model in which
the gas accretion moves forward in lock-step with dark-matter
collapse.

\begin{figure}[t] 
\epsscale{1.22}
\plotone{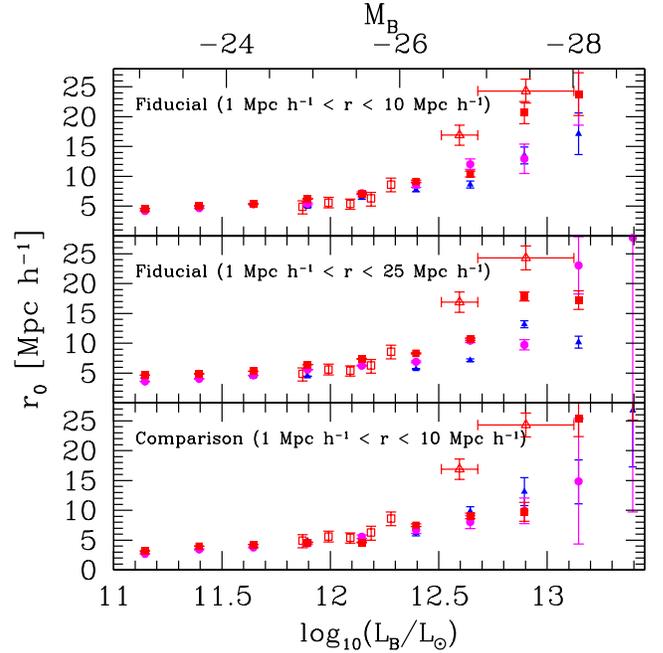}
\caption{{\em Top:} Correlation length as a function of luminosity and
redshift. The closed triangles, circles, and squares correspond to
redshift  bins $z$ = 1.5, 2.0, and 3.0 respectively, with 1-sigma Poisson error bars.
Open squares
correspond to the $\gamma=1.8$ fits of Porciani \&  Norberg (2006) to
the 2QZ survey at redshifts  $z$ = 0.93, 1.19, 1.41, 1.60, 1.79, and 1.98, 
with the mean $M_B$ being taken from  Croom  \etal (2005). Open triangles
correspond to the Shen \etal (2007) SDSS data  points for $2.9<z<3.5$ and
$z>3.5$, but are fitted for $\gamma=2.0$ rather  than $1.8$ with
i-band to B-band conversion taken from Hao \etal (2005).} 
\label{fig:r0}
\end{figure}

A second way to quantify our results is by using the correlation
length, $r_0,$ the scale at which $\xi_{\rm qq} = 1.$  
If $\xi_{\rm qq} \propto r^{-\gamma},$ this occurs at
$r_0^\gamma = \xi_{\rm qq}(r) r^\gamma$ for all choices of $r$.  Thus we can 
compute $r_0$ by averaging this quantity over all radial bins $r_k$ from
1 to 10 $h^{-1}$ Mpc
\be 
r_0(L,z) = \left[ \frac{\sum_k
w(r_k,L,z) \xi_{\rm qq}(r_k,L,z) r_k^{\gamma}}  {\sum_k w(r_k,L,z)} \right]^{1/\gamma},
\label{eq:r0}
\ee where we choose to set $\gamma = 1.8$ and $\xi_{\rm qq}$  is the 
quasar
auto-correlation function as a function  of luminosity $L$ and
redshift $z$.  In Figure \ref{fig:r0} we compare the results of this
analysis with correlation length measurements from the 2QZ (range
$1<z<2$) and SDSS (range $3<z<5$) surveys. A selected list of the 
computed correlation lengths is also given in Table \ref{table1}.

The fiducial model, shown in the top panel, indicates that  below
$L_B\approx 10^{12} L_\odot$ there is little  evolution in the
correlation length either as a function of redshift or luminosity. The
calculated correlation lengths are also in good  agreement with the
Croom \etal (2005) results.  Above $L_B\approx 10^{12} L_\odot,$ on
the other hand,  the variability of the correlation with luminosity is
more evident, however even this seems to fall slightly short of the very 
large
correlation length seen in the SDSS sample (although our quoted Poisson 
errors are smaller than the systematic errors from the binning 
procedure) . We note that our 
our highest redshift bin ($z=2.25-4.0$) has a
mean redshift less than that of the  $z>3.5$ SDSS bin, but the lower 
$2.9<z<3.5$ SDSS bin is comparable to our $z=3$ predictions.
The primary difference between results probably stems from the fact that
the SDSS
measurement considers pairs  with separations from $4 \,  h^{-1}
{\rm Mpc} < r < 150 \, h^{-1} {\rm Mpc}$, which is a vastly larger  range
than  can be probed with current simulations that retain high
resolution in individual galaxies. In fact, it could be reasonably
argued  that a  box of 1000 $h^{-1}$ Mpc is necessary to study
correlations on this  scale.  Additionally, considering correlations
to such  large radii is fraught with potential difficulties, since for
redshifts $z \lesssim 1$  the galaxy-galaxy correlation function
is expected to steepen  for separations larger than $60 h^{-1}$ Mpc (see
Springel, Frenk \& White 2006). Were AGN/quasars to more  closely trace 
the
underlying dark matter auto-correlation function than the normal
galaxy population at high redshift then the effective power law for the
auto-correlation function would be expected  to change  by almost 50\%
as one moves from 5 $h^{-1}$ Mpc to 20 $h^{-1}$ Mpc (Porciani \&
Norberg 2006). Further, fitting  steeper power laws will inherently
tend to produce larger correlation lengths.

The steepening of the galaxy-galaxy correlation
function is a  product of the underlying bias of the galaxy population
and  the  transition to the non-linear regime in the dark matter power
spectrum,  which occurs at $k>k_{nl}\simeq 0.1 \, h$ Mpc${}^{-1}$ at low
redshift. This  scale evolves comparatively slowly until $z\approx 1$,
above which it begins  to recede quickly, and by $z\simeq3$ we find
$k_{nl}\simeq 0.3 \, h$ Mpc${}^{-1}$.
Semi-analytic models of the
galaxy-galaxy correlation function are able  to roughly preserve the
location of this steepening point (see, {\em e.g.,}  Springel  \etal
2006) but do so through a  rapidly increasing bias with redshift. If
the bias of the AGN population  does not increase sufficiently quickly
with redshift then the clustering  statistics will directly measure
the underlying evolution in the  non-linear scale. It is worth noting
that current surveys at $z\approx3$  definitely straddle the turnover
in the dark matter correlation  function.
Thus the enhanced correlation 
seen in the SDSS might be related to a change in the radial 
position of the steepening of the correlation function with redshift. 
The precise details are dependent 
upon the redshift evolution of the bias of the AGN population and, as we 
have indicated, our 
simulation box is too small to make any firm statements.

However, to study the dependence of our results on the much smaller distances
we can probe,  we recalculated eq.\ (\ref{eq:r0}) using a range
of  separations from  $r_k = 1$ $h^{-1}$ Mpc to  25 $h^{-1}$
Mpc.  These values, shown in the second panel of Figure \ref{fig:r0}
demonstrate the correlation is decreased systematically when one
includes more information from large separations.  As discussed above
in relation to Figure \ref{fig:xi}, the origin of these differences is
most likely to be the excess contribution at small separations from
the one-halo term, which becomes less important as moves from the
shorter 10 $h^{-1}$ Mpc cutoff to the longer 25 $h^{-1}$ Mpc cutoff.
Even this modest change in the outer cutoff can change the
correlation lengths by as much as 40\% (specifically in the z=1.5, 
$L_B=12.4 \;L_\odot$ bin) although the mean change is close 
to 15\%. 

Finally,  in the lower panel, we show the results from our comparison
simulation, which again displays  similar trends as in the AGN
feedback run, but with a lower level of clustering. For all
models and separations it is difficult to draw conclusions about the
redshift evolution  of $r_0$.  Indeed, our results support
the idea of using correlation function evolution models that are
roughly constant in comoving coordinates since we see little evolution 
in our comparison models (for example) at $\log_{10}(L_B/L_{\rm 
bol})<12.5$.

\section{X-ray Selected AGN}

X-ray selection is widely believed to be an unbiased method for
selecting AGN candidates, which is largely free from the
obscuration and incompleteness issues that affect optical catalogs 
(\eg Yang \etal 2006).  While the exact nature of the optical versus X-ray
light curves is the subject of debate (as summarized in Hopkins \etal
2008, and  references therein),  we examine the X-ray clustering of
our catalog based on the same assumptions as our optical
catalog. In this case the only  differences between  X-ray and optical
selection come from bolometric correction factors, as even
obscuration of optical systems should not impact the overall
correlation unless obscuration is somehow a function of position.
Additionally, so long as the average  lifetime of the X-ray
bright period is not longer than $\approx 1$ Gyr, and hence peculiar
velocities do not lead to systems moving an appreciable  distance, our
lifetime assumptions should not have a significant impact on
clustering.  Of course this is not true for the luminosity
function,  which is very sensitive to changes in lifetimes, and    we
have previously shown (TSC06) that the hard X-ray luminosity  function
calculated  from our model reproduces the observations of Ueda \etal
(2003).

To calculate the ratio of the intrinsic X-ray luminosity, $L_X$, to the
bolometric luminosity, $L_{\rm Bol}$,  within our simulation we  use
the following two-polynomial fits from Marconi \etal (2004),
\[
\log[L_{\rm Bol}/L_X(2-10\; {\rm keV})]=1.54+0.24{\mathcal
L}+0.012{\mathcal L}^2+0.0015{\mathcal L}^3,
\]
\begin{equation}
\log[L_{\rm Bol}/L_X(0.5-2\; {\rm keV})]=1.65+0.22{\mathcal
L}+0.012{\mathcal L}^2+0.0015{\mathcal L}^3,
\end{equation}
where ${\mathcal L}=\log(L_{\rm Bol})-12$, and $L_{\rm Bol}$ is given
in $L_\odot$. 
While a large number of  estimates for bias and correlation lengths
exist for $z\lesssim1$ (\eg  Mullis \etal 2004; Gilli \etal 2005;
Basilakos \etal 2004, 2005; Yang \etal 2006; Miyaji \etal 2007), the
current observational data is  too sparse at redshifts $z>2$ to
provide reliable statistics  on redshift evolution.  However,
cross-correlating luminous blue galaxies  and AGN allows a calculation
of the bias of the AGN population at  $z\simeq3$ (Francke \etal
2008). We therefore calculate both the bias  and correlation
functions of our simulation, but split the catalog into only  two 
redshift bins: $z=1.2-2.0,$ which we label as $z=1.75,$ and  $z=2.0-4.0,$
which we  label as $z=3.0$.
Within these bins we then calculate
correlation functions and bias using the procedures outlined in
section 3.2.

\begin{figure}
\epsscale{1.2} 
\plotone{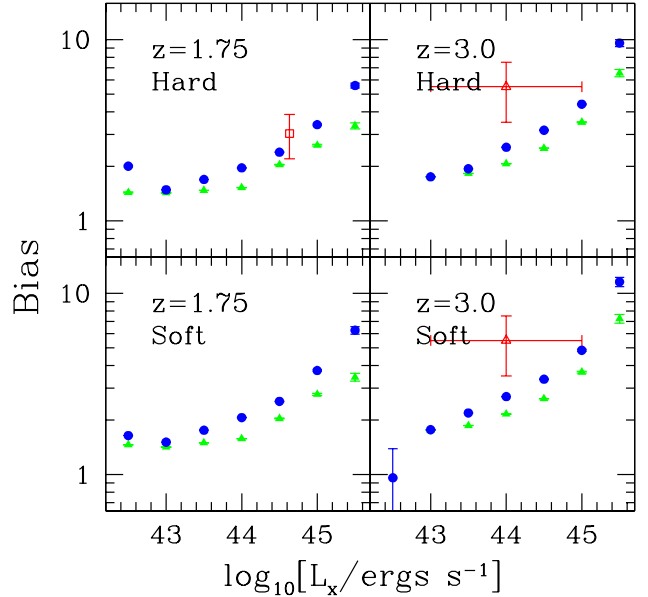}
\caption{Bias of X-ray selected AGNs as a function of  redshift and
X-ray luminosity. The green triangles are from the  comparison run,
while the blue circles are from the fiducial  feedback model,
both with 1-sigma Poisson error bars. The open
square data point on the $z=1.75$ plot is from Yang  \etal (2006), and
corresponds to the bias for their $z=1.5-3.0$ bin (no variance for the
mean of the luminosity bin is given). The open triangle is from
Francke  \etal (2008) and gives their AGN bias calculated from the
cross-correlation  function of AGN and luminous blue galaxies at
$z\simeq3$. The error bars  encapsulate their range in luminosity and
the mean is likely rightward of  the central value. While our fiducial
run is in good agreement with the  Yang \etal (2006) results, the
$z\simeq3$ result is clearly lower than  Francke \etal (2008) data.}
\label{fig:biasX} 
\end{figure}  

In Figure  \ref{fig:biasX} we plot the bias of our catalog for both
hard and soft  bands in our two redshift bins. As for the optical
catalog, we find  that the simulation with quasar feedback has a
higher bias than the comparison run.  Again, the primary origin of
this difference is that feedback forces AGN of a given luminosity into
more massive, and thus more biased halos.  The fiducial run hard
X-ray data in the $z=1.75$ bin are a good fit to the Yang \etal (2006)
bias estimate from the CDF-N and CLASXS fields, which suggests that
including  feedback is necessary to   bias halos sufficiently to match
observations.  Furthermore, 
comparison to the $z=3$ bias estimates of Francke \etal
(2008), calculated using the cross-correlation of AGN and  luminous
blue galaxies, shows that our numbers are low relative to these
observations, and that AGN feedback is absolutely necessary to match
these  data. We achieve a marginal  agreement if we take  the
low value  of their error and also the higher end of the luminosity
range, which is  plausible since we plot the center of the range and
the the mean of their bin is likely rightward of the central value.
It is also worth noting that these results show that in our model the
luminosity  dependence begins to become more noticeable above
$10^{43}$ erg s${}^{-1}$. 

Before examining our results further we mention that since the 
luminosities of the X-ray data we are considering are considerably lower 
than the equivalent optically-selected catalog, the observations we 
consider may well be probing  different AGN fueling 
mechanisms as compared to our major merger model. In particular Hopkins 
\& Hernquist (2006) have suggested that the limiting upper luminosity 
for 
where secular (\ie Seyfert) effects or minor mergers become important is 
around 
$M_B\simeq -22$. This corresponds to an X-ray luminosity of $10^{44}$ 
erg s${}^{-1}$, which is roughly in the middle of our considered X-ray 
luminosity range. However, since we compare directly to observations 
with mean luminosities above $10^{44}$ 
erg s${}^{-1}$ we can be reasonably confident that major mergers are the 
dominant physical process in these systems.

Both the  fiducial and comparison runs correspond to the same bias at
the faint  end, ignoring the faintest bins at  $z=3$ which are at the
limits of our resolution. However, the overall sensitivity to
luminosity is higher for the fiducial run when examined over
the entire range  $L_X=[10^{43},10^{45.5}]$ erg s${}^{-1},$
showing an increase in the
bias of roughly a  factor of  4 in the hard and soft X-ray bands at
$z=1.75$, and a factor of close to  $6$ in the $z=3.0$ bin.  Hints of
this dependence are observed in the optical catalog over  the range
$\log(L_B)=[11.8,12.6]$, although it is difficult to determine
visually  since  the brighter bins have large error bars while the
faint-end cutoff  at $L_B=10^{11.4} L_\odot$ does not probe as low in
luminosity as the X-ray  catalog. In the X-ray data, the ratio of the
hard X-ray bias  of the fiducial run to the  comparison run is
{$\approx 1.6$} for $L_X=10^{45.6}$ erg s${}^{-1}$ at  both $z=1.75$
and  $z=3.0$ and the soft X-ray numbers are similar.  The
$L_X=10^{45.5}$ erg s${}^{-1}$ bin also shows redshift
evolution, with its bias decreasing by a factor of two from  $z=3$ to
$z=1.75$.

\begin{figure}[t]
\epsscale{1.21}
\plotone{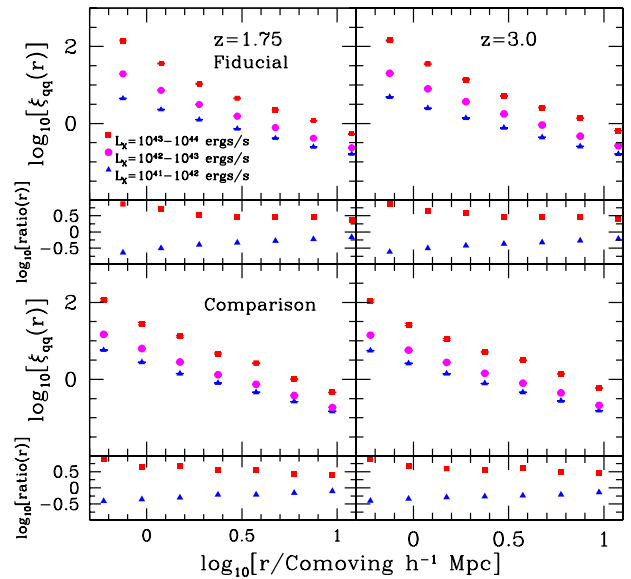}
\caption{Correlation function of X-ray selected AGNs. We have divided 
the simulation into two redshift ranges, $z=2.25-4.0$, which we label 
$z=3,$ and$ z=1.1-2.25$ which we label $z=1.75$. We then bin into 3 decades 
of $L_X$, ranging from $10^{41}$ erg s${}^{-1}$, to $10^{44}$ erg 
s${}^{-1}$. The ratio of the $3.2\times10^{43}$ erg s${}^{-1}$ to
$3.2\times10^{44}$ erg  s${}^{-1}$ X-ray correlation functions
increases by a factor of three in the  one-halo regime, with a similar 
rise being observed in the optically selected catalogs.
All error bars are 1-sigma Poisson estimates.
}
\label{fig:xiX}
\end{figure}

In Figure \ref{fig:xiX} we plot the correlation function of our
X-ray selected AGN.
While our estimates seem to be in good agreement  with the data (the
little studied $z=3$ bin aside),  the correlation lengths in this
plot appear to be smaller than the observed data.  
Table \ref{table1} quantifies the correlation lengths extracted from the X-ray
data using eq.\ (\ref{eq:r0}), as well as selected correlation lengths and
biases from throughout this paper.   Observationally, Yang \etal (2006) give a
combined CDF-N, CLASXS X-ray correlation length of
$r_0=6.1^{+0.4}_{-1.0}h^{-1}$ Mpc, albeit with a shallow $\gamma=1.47$
slope.  Matching this correlation length at $z=1.75$ with our
luminosity  binned data  requires a mean luminosity greater than
$3.2\times 10^{44}$ erg s${}^{-1}$. However, just like the optical data 
these surveys have a much higher cutoff radius than our simulation. For 
example the CLASXS field discussed in Yang \etal (2006) considers pairs 
with separations up to 200 $h^{-1}$ Mpc, well beyond
beyond the radius at which the down-turn occurs
in the galaxy-galaxy correlation function.

Recently, using luminosity binning, Plionis
\etal (2008) have suggested that the wide variation in the correlation
lengths of the CDF-S and CDF-N  can be reconciled. 
They find evidence  of strong  evolution in clustering
as a function of luminosity, with the  correlation length in the hard
X-ray band increasing from $\approx
6  h^{-1}$ Mpc to $\approx 18 h^{-1}$ Mpc, with a $0.7$ dex increase
in luminosity. A comparison to our X-ray correlation  lengths
in Table \ref{table1} shows that even our fiducial ``light bulb'' model
cannot produce this level of luminosity  dependence,  nor can
we produce the  same underlying clustering. Thus although we do find a 
somewhat stronger trend for $r_0$ to increase with
$L_X$ than $L_B$, our results are at odds with Plionis et al. (2008)
 It is also 
worth noting
that their quoted correlation length for the highest flux in the soft
band  of $\approx 30$ $h^{-1}$ Mpc is considerably larger than that
quoted for  the IRAC Shallow Cluster Survey (Brodwin \etal 2007),
$r_0=19.14^{+5.65}_{-4.56} h^{-1}$ Mpc at $z=0.97$. This implies 
that the comparatively low luminosity AGN (mean $L_X \approx 10^{43}$ erg 
s${}^{-1}$) they sample 
are more strongly clustered at high redshift than $z=1$ galaxy clusters.
Ultimately more clustering data is needed to help understand the high 
redshift clustering of X-ray selected AGN, and we eagerly anticipate 
future all-sky surveys.

\section{Discussion}

We have presented an analysis of our simulated quasar/AGN catalog,
focusing on   the dependence of real-space clustering on redshift,
luminosity, and  selection. Our model does not follow the detailed
accretion history  onto the central supermassive black hole (\eg
Hopkins \etal 2005a, 2005b, 2005c, 2006a, 2006b, 2007a), but rather
takes  a simple one-to-one correspondence between black hole mass and
luminosity.  Nonetheless we capture much of the essential  physics in
AGN formation and feedback.  As a consistency check on our earlier
work, we showed that a model that does not  include feedback precisely
follows the predicted luminosity function of the Wyithe \& Loeb (2003)
model, and thus fails badly at low redshift  by overpredicting the
observed numbers counts.

On the other hand, the qualitative luminosity function behavior is
reproduced (TSC06) when feedback is included. Our clustering results
are also in broad agreement with observed data, the main difference
being somewhat less evolution with redshift  than observed and
somewhat smaller correlations lengths, although our  bias values are
in quantitative agreement within the observational errors.
Furthermore for our ``light bulb''  model,  the dependence  of
clustering with luminosity is weak at the luminosities probed by
current surveys.  Although the underlying relationship between quasar
luminosity and black hole mass is likely to more complex than the
simplified model assumed in this study (\eg Ganguly  \etal 2007),
modeling these complexities dose not appear to be necessary to
understanding current clustering measurements.

While the assumptions used in our calculations limit our analysis to
scales above $\approx 0.5$  Mpc $h^{-1}$, we are still able to
clearly observe one-halo effects, which occur within $\approx 2$ 
Mpc $h^{-1}$ for luminous AGN.  Significantly, the luminosity
dependence is more visible in this part of the correlation function as
the two systems are embedded in a more highly biased halo.  For
example, the ratio of the $3.2\times10^{43}$ erg s${}^{-1}$ to
$3.2\times10^{44}$ erg  s${}^{-1}$ X-ray correlation functions
increases by a factor of three in the  one-halo regime.  Deep quasar
pair data would thus be extremely useful in helping to determine
luminosity dependence in more detail.

However, we reemphasize that a straightforward comparison of current
observations to our results, or those of any simulations, is not
possible. While larger samples have made observational studies more
robust, there still remain differences in fitted scales for
correlation functions and the assumed slope of power-law fits. This is
particularly important in the context of calculating correlation
lengths, as departures from power-law fits occur both on small scales
and large scales. On small scales the  one-halo term produces a
steepening in the effective index, while on large scales the
transition from the non-linear to linear regime in the dark-matter
power spectrum also produces a steepening of the effective index. We 
also note that redshift-space distortions (\eg Croom et al. 2005; da 
Angela \etal 2005, 2008), can also produce departures from pure 
power-law behavior. Ultimately,
the true power-law slope observed for the AGN/quasar population  will
depend on the underlying bias.

Given these facts and our modest outer radius of 25 $h^{-1}$ Mpc, the
fits we derive should be treated with due caution. For  example, even
for modest changes  in the outer cutoff from 25  $h^{-1}$ Mpc to 10
$h^{-1}$ Mpc, our correlation lengths  can increase by
as much as 40\%, although the mean change is close to 15\%.  The
increase is directly associated with the one-halo contribution
being given more weight in the case
with the shorted outer cutoff, although we emphasize that all
fits below 10 $h^{-1}$ Mpc are within the non-linear scale at the
epochs we are considering.

The redshift and luminosity dependence of large-scale clustering is a
product of two competing effects: growth of the non-linear mass scale
with time  and a decrease in the mass scale of the
quenching threshold that limits the supply of fuel to AGN.  This
quenching is primarily a function of the mean density of the gas, which
controls its cooling rate.  Thus at high redshifts, when radiative
cooling is extremely efficient, feedback is weak, and the
luminosity of black holes grows along with the nonlinear mass scale.
However, once feedback  is able to heat gas to a cooling time longer
than the Hubble time, the fuel supply for luminous  AGN is quenched,
and this quenching become more efficient as the universe expands.
The net result is a peak in AGN activity  at redshift $z\approx 2$:
AGN formed after this redshift correspond to low-mass,
low-bias halos and show a weak luminosity dependence, and AGN formed
before this redshift correspond to a wide range of biases and show an
appreciable luminosity dependence.  Thus the clustering behavior of
AGN is a direct result both of  the evolution of dark
matter halos and the physics of AGN feedback.

\acknowledgments

We are grateful to Michael Strauss for useful comments and Phil 
Hopkins for making his compilation of quasar luminosity 
function data freely available. We also thank an anonymous referee for 
suggestions that improved the content of the paper. Figure 1 was 
produced using the ``Splotch'' package ({\tt 
http://dipastro.pd.astro.it/}\~{\tt cosmo/Splotch/}).
Computing was performed at WestGrid (under a RAC grant), ACEnet and on the {\em 
Computational Astrophysics Laboratory} at Saint Mary's University.  R.J.T.\
acknowledges funding via a Discovery  Grant from NSERC, the Canada
Research Chairs program and the Canada Foundation for
Innovation. M.R. was funded by an NSERC USRA. HMPC acknowledges funding   
from NSERC 
and the support 
of
the Canadian Institute for Advanced Research.

\fontsize{10}{10pt}\selectfont

\end{document}